\documentclass[a4paper,10pt,twocolumn]{article}

\usepackage[english]{babel}
\usepackage[utf8]{inputenc}
\usepackage[T1]{fontenc}

\usepackage[top=1.5cm, left=1.5cm, right=1.5cm, bottom=1.5cm]{geometry}

\renewenvironment{abstract}{\bf\small {\em\ Abstract---}}{}

\usepackage{amsfonts,amssymb,amsmath,amsthm}
\usepackage{subfigure}
\usepackage{graphicx}
\usepackage[footnotesize]{caption}

\usepackage{amsfonts,amssymb,amsmath,amsthm}
\newtheorem{theorem}{Theorem}[section]

\DeclareMathOperator*{\argmin}{arg\,min}
\title{Single molecule localization by $\ell_2-\ell_0$ constrained optimization}

\author{ Arne Bechensteen$^1$,  Laure Blanc-Féraud$^1$ and Gilles Aubert$^{2}$.\\
  \footnotesize $^1$Université Côte d'Azur, CNRS, INRIA, Laboratoire I3S UMR 7271, 06903 Sophia Antipolis, France.\\  \footnotesize $^2$Université Côte d'Azur, UNS, Laboratoire J. A. Dieudonn\'{e} UMR 7351, 06100 Nice, France.\ } \date{\empty} 

\begin{document}

\maketitle

\begin{abstract} 
Single Molecule Localization Microscopy (SMLM) enables the acquisition of high-resolution images by  alternating between activation of a sparse subset of  fluorescent molecules present in a sample and localization. In this work, the localization problem is formulated as a constrained sparse approximation problem which is resolved by rewriting the $\ell_0$ pseudo-norm using an auxiliary term. In the preliminary experiments with the simulated ISBI datasets the algorithm yields as good results as the state-of-the-art in high density molecule localization algorithms.  
\end{abstract}

\section{Introduction}
\label{sec:introduction}

The exploration of fluorescent molecules in microscopy has made it possible to bypass the limit of resolution imposed by the diffraction limit and obtain images often referred to as super-resolved images (see \cite{hess_ultra-high_2006}, \cite{betzig_imaging_2006} and \cite{rust_sub-diffraction-limit_2006}).  The principle of SMLM is to excite a sparse number of fluorescent molecules for each acquisition and locate each molecule with an algorithm. Since fluorescence microscopy can be used for live imaging, the subject may  move during the acquisitions yielding a faulty reconstruction. More molecules excited for each single acquisition will lower the total acquisition time. Therefore the subject has less time to move with high-density acquisition. This demands an efficient reconstruction as more than one molecule could be present in the same diffraction disk. 

In this work we aim to reconstruct precisely high-density 2D images. We model the localization problem as a $\ell_2-\ell_0$ constrained minimization problem. A similar approach, the $\ell_2-\ell_0$ penalized minimization problem, has been previously studied with success \cite{gazagnes_high_2017}. However, a common problem with penalized regularizations terms is to choose the trade-off parameter between the data fidelity term and the regularization term. In this paper we look at the $\ell_2-\ell_0$ \textit{constrained} version as the sparsity parameter is easier to handle. We propose a  minimization algorithm based on exact reformulation of the $\ell_0$ pseudo-norm by introducing an auxiliary variable. 

\section{Acquisition system modelling}
\label{sec:first-section}
The fluorescence molecules are observed through an optical system and thus the signal, $x\in \mathbb{R}^{ML\times ML}$, is diffracted. This is modeled by the Point Spread Function (PSF) which is convolved with the signal. This operation is denoted $H:\mathbb{R}^{ML \times ML}\rightarrow \mathbb{R}^{ML \times ML}$, and we have chosen to use the Gaussian approximation of the PSF

\begin{equation}
    PSF(x_1,x_2)=\frac{1}{\sqrt{2\pi}\sigma_s}\exp\left[ -\frac{x_1^2+x_2^2}{2\sigma_s^2}\right]
\end{equation}
where $\sigma_s$ is spatial standard deviation. 

A sensor captures the convolved signal in a lower resolution which is modeled by a reduction operator $R_L:\mathbb{R}^{ML \times ML}\rightarrow \mathbb{R}^{M \times M}$ which is defined as
\begin{equation}
    R_L(x)=RXR^T
\end{equation}
where  $X\in\mathbb{R}^{ML}\times\mathbb{R}^{ML}$ and $X$ is the vector $x$ arranged as a matrix.  $R$ is a matrix of $\mathbb{R}^{M}\times \mathbb{R}^{ML}$ with $L$ 1's in each line and $R^T$ is the transposed matrix of $R$.

During this image acquisition the signal is corrupted by different kind of noise $\eta$. In this work we assume the noise to be Gaussian additive noise. The system can therefore by modeled as
\begin{equation}
    d=R_L(H(x))+\eta
\end{equation}
with $d\in\mathbb{R}^{M \times M}$ and $x\in\mathbb{R}^{ML \times ML}$. Further on we will refer the linear operation $R_L(H(x))=Ax$ to ease the notations. 

The localization is done on a finer grid $x\in\mathbb{R}^{ML \times ML}, L>1,$ than the observed signal $d\in \mathbb{R}^{M \times M}$. Therefore, the inverse problem is underdetermined. We include a sparse constraint term as only a few number of molecules are excited for each acquisition, and we note the maximum number of molecules we want to reconstruct as $k$ which is the sparsity parameter. This constraint is introduced as $\iota_{\|\cdot\|_0\leq k}(x)$, where  $\|x\|_0=\#\{x_i , i=1,\cdots N : x_i\neq 0\}$, and will be, by abuse of language, referred to as the $\ell_0$ norm.  $\iota_C(x)$ is the indicator function such that 
$$
\iota_C(x) =\begin{cases} 0 \text{ if } x\in C\\
+\infty \text{ otherwise }
\end{cases}
$$ 
Furthermore, we add the constraint that each reconstructed molecule must have a positive value since we reconstruct their intensity.
We search therefore
\begin{equation}
       \hat{x}\in \argmin_{x\in\mathbb{R}^{ML\times ML }} \frac{1}{2}\|Ax-d\|^2+\iota_{\|\cdot\|_0\leq k}(x)+\iota_{\cdot\geq 0}(x)
    \label{eq:trueq}
\end{equation}

This problem is non-convex and non-continuous as well as NP-hard due to the nature of the $\ell_0$ norm. The problem has been extensively studied, and among the approaches to ease or  resolve the problem we find relaxations of the $\ell_0$ norm \cite{soubies_unified_2017} and greedy algorithms \cite{soussen_bernoulli_2011}. In this paper we use an exact reformulation of the $\ell_0$ norm that  we will present in the next section.

\section{Exact reformulation of the $\ell_0$ norm}
\label{sec:second-section}
The article \cite{yuan_sparsity_2016} inspired us to extend their results to the $\ell_2-\ell_0$ constrained problem. 
They propose to rewrite the $\ell_0$ norm as a convex minimization problem by introducing an auxiliary variable $u$. 
\begin{equation}
    \|x\|_0 = \min_{-1\leq u \leq 1} \|u\|_1 \text{ s.t } \|x\|_1= <u,x>
\end{equation}

With the reformulation of the $\ell_0$ norm we can rewrite our problem as
$$
\min_{x,u} \frac{1}{2}\|Ax-d\|+\iota_{\cdot\geq 0}(x)+ I(u) \text{ s.t. } \|x\|_1=<x,u>
$$
where $I(u)$ is :
$$
I(u)=\begin{cases}
0 \text{ if }\|u\|_1 \leq k \text{ and } -1\leq u \leq 1\\
\infty \text{ otherwise }
\end{cases}
$$

We can then define a new penalty term 
\begin{equation}\label{eq:Grho}
G_\rho(x,u)= \frac{1}{2}\|Ax-d\|^2 +\iota_{\cdot\geq 0}(x)+I(u)+\rho(\|x\|_1-<x,u>)
\end{equation}
$\rho$ is a trade-off penalty to ensure that the equality constraint between the $x$ and the $u$ variables is verified. 
\begin{theorem}\label{the:tho}
Assuming that $A$ is full rank, the penalized functional $G_\rho(x,u)$ (\ref{eq:Grho}) has the same local and global minimizers as the constrained initial problem (\ref{eq:trueq})  when $\rho>\|A^Td\|_2\left(\frac{2\sigma_1(A)^2}{\sigma_2(A)^2}+1\right)$. $\sigma_1(A)$ and $\sigma_2(A)$ represent the largest and smallest singular value of the matrix $A$, respectively. 
\end{theorem}
The general idea  to resolve $G_\rho(x,u)$ is to solve the problem $[\hat{x}^0,\hat{u}^0]=\argmin G_{\rho^0} (x,u)$  with a $\rho^0$ small as the non convexity comes form the scalar product $<x,u>$.   Then we increase the $\rho$ for each iteration, and resolve $[\hat{x}^{n+1},\hat{u}^{n+1}]=\argmin G_{\rho^n} (\hat{x}^n,\hat{u}^n)$. This will hopefully give a good initialization for the final minimization, that is when $\rho$ is  according to theorem \ref{the:tho}.

 The minimization of $G_{\rho} (x,u)$ is done by using the Proximal Alternating  Minimization algorithm (PAM) \cite{attouch_proximal_2008} which ensures convergence to a critical point. The problem is biconvex and  we alternate between minimization with respect to $x$ and with respect to $u$. 
 
\textbf{x-step}: The minimization with respect to $x$ using the PAM algorithm is 
\begin{align*}
 x^{n+1}=\argmin_x \frac{1}{2}\|Ax-d\|^2 &+ \rho(\|x\|_1 - <x,u^n>)\\
 &+\iota_{\cdot\geq 0}(x) +\frac{1}{2c^n}\|x-x^n\|_2^2   
\end{align*}

where $c^n>0$, and the above problem can be solved using classical minimization schemes such as FISTA \cite{beck_fast_2009}.

\textbf{u-step}: The second step is to minimize with respect to $u$ the following problem
$$
u^{n+1}=\argmin_{-1\leq u \leq 1} \frac{1}{2b^n}\|u-u^n\|_2^2-\rho<x^{n+1},u> \text{ s.t. } \|u\|_1 \leq k
$$
where $b^n>0$. The above problem is  equivalent to 
$$
u^{n+1}=\argmin_{-1\leq u \leq 1} \frac{1}{2}\|u-(u^n+\rho b^nx^{n+1})\|^2 \text{ s.t. } \|u\|_1 \leq k
$$
For simplicity we denote $z=u^n+\rho b^nx^{n+1}$. 
Since this is a symmetric problem, we can rewrite it as 
$$
|u^{n+1}|=\argmin_{0\leq u \leq 1} \frac{1}{2}\|u-|z|\|^2 \text{ s.t. } \|u\|_1 \leq k
$$
where $u^{n+1}$ can be reconstructed with $sign(z)|u^{n+1}|$.
This minimization problem is a variant of the  knapsack problem  which can be resolved using classical minimization schemes such as \cite{doi:10.1155/S168712000402009X} which we used in our algorithm.
\begin{align*}
|u^{n+1}|= \argmin_{0\leq u \leq 1} &\frac{1}{2}<u,u>-<u,|z|> \\&\text{ s.t. } \left(\sum_i u_i \right)\leq k
\end{align*}

\section{Results}
\label{sec:third-section}

We compare our method to the $CEL0$-minimization from \cite{gazagnes_high_2017} which is based on the $\ell_2-\ell_0$ penalized problem. The algorithms are tested on a simulated dataset accessible from the ISBI-2013 challenge \cite{sage_quantitative_2015}.  The dataset is of 8 tubes of 30 nm diameter, where the acquisition is simulated with a size of $64\times 64$ pixels where each pixel is of  size 100nm, and the PSF is modeled by a Gaussian function with a Full Width at Half Maximum (FWHM) equals to 258.21nm. In order to test our algorithm on high-density acquisitions, we have summed 5 images at a time, simulating a total number of 72 images. We localize the molecules on a $256\times 256$ pixel grid, where each pixel has the size of 25nm, and the center of the detected pixels are used to estimate their position in nm. This translates to finding an $x\in \mathbb{R}^{ML\times ML}$ from an acquisition $d\in \mathbb{R}^{M\times M}$ with $M=64$ and $L=4$.  We set $\rho^0=10^{-4}$ and $k=170$ after a few numerical tests on a single acquisition. In the case of $CEL0$ we choose the regularization parameter $\lambda=0.23$ such that on average the algorithm reconstructs 170 molecules for each image. 

\begin{figure}[]
    \centering
    \includegraphics[width=0.5\textwidth]{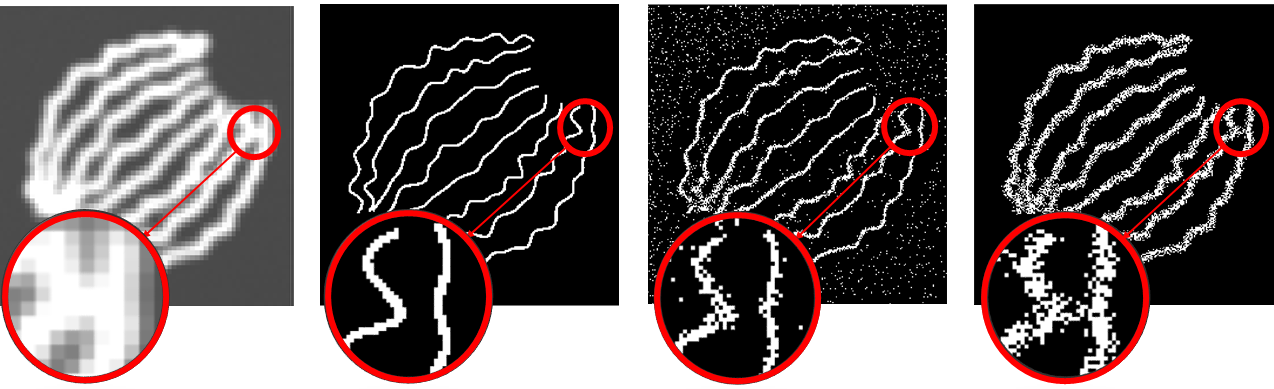}
    \caption{Reconstruction results. From left to right: Observed image, ground truth, $CEL0$ and the proposed algorithm reconstruction. }
    \label{fig:comp}
\end{figure}

The performance is evaluated with the Java tool obtained from the ISBI-SMLM site and we consider here the Jaccard index as a measure of performance. The Jaccard index is the ratio between the correctly reconstructed molecules and the sum of correctly reconstructed-, true positives- and false positives (FP) molecules.  In Table \ref{tab:res} we observe that for a  tolerance disk of the Jaccard index of 50 nm the $CEL0$ method \cite{gazagnes_high_2017} reconstructs better. However, when increasing the tolerance we observe that our proposed method reconstructs the molecules more precisely.  In Figure \ref{fig:comp} we observe that the algorithm $CEL0$ distinguishes two close tubes and for our proposed model this is less clear. We observe that the  $CEL0$ reconstruct many FP in comparison to our proposed method. Note that a greater $\lambda$ would remove FP but also reconstruct less molecules. 

\begin{table}[]
\centering

\begin{tabular}{|l|l|l|l|l|l|}
\hline
                        & \multicolumn{5}{l|}{Jaccard index (Ji) (\%)}                                       \\ \hline
Method - Tolerance (nm) & 50            & 100           & 150           & 200           & 250           \\ \hline
Proposed model          & 10.1 & \textbf{14.7} & \textbf{15.1} & \textbf{15.1} & \textbf{15.1} \\ \hline
IRL1-CEL0               & \textbf{11.6}         & 12.9         & 13.1          & 13.2              & 13.3             \\ \hline
\end{tabular}
\caption{The Jaccard index obtained for the two methods and the tolerance disk.}
\label{tab:res}
\end{table}

\section{Conclusion}
\label{sec:conclusion}

In this paper we have addressed the problem of high-density super-resolution imaging. We have modeled the acquisition system as a $\ell_2-\ell_0$ \textit{constrained} problem and proposed a minimizing algorithm. The first numerical results are good but the computational time is quite important due to the minimization of $G_{\rho^n}$ for several $\rho^n$  and we are working on other approaches for the optimization of the sparse constrained problem.



\begin{thebibliography}{10}
\bibitem{attouch_proximal_2008}
H. Attouch, J. Bolte, P. Redont, and A. Soubeyran.
\newblock Proximal alternating minimization and projection methods for
  nonconvex problems. {An} approach based on the {Kurdyka}-{Lojasiewicz}
  inequality.
\newblock {\em arXiv:0801.1780 [math]}, January 2008.
\newblock arXiv: 0801.1780.

\bibitem{beck_fast_2009}
A.~Beck and M.~Teboulle.
\newblock A {Fast} {Iterative} {Shrinkage}-{Thresholding} {Algorithm} for
  {Linear} {Inverse} {Problems}.
\newblock {\em SIAM Journal on Imaging Sciences}, 2(1):183--202, January 2009.

\bibitem{betzig_imaging_2006}
E. Betzig, G. H. Patterson, R. Sougrat, O.~W. Lindwasser, S.
  Olenych, J.S. Bonifacino, M.~W. Davidson, J.
  Lippincott-Schwartz, and H.~F. Hess.
\newblock Imaging {Intracellular} {Fluorescent} {Proteins} at {Nanometer}
  {Resolution}.
\newblock {\em Science}, 313(5793):1642--1645, September 2006.

\bibitem{gazagnes_high_2017}
S.~Gazagnes, E.~Soubies, and L.~Blanc-Féraud.
\newblock High density molecule localization for super-resolution microscopy
  using {CEL}0 based sparse approximation.
\newblock In {\em 2017 {IEEE} 14th {International} {Symposium} on {Biomedical}
  {Imaging} ({ISBI} 2017)}, pages 28--31, April 2017.

\bibitem{hess_ultra-high_2006}
S.~T. Hess, T. P.~K. Girirajan, and M.~D. Mason.
\newblock Ultra-{High} {Resolution} {Imaging} by {Fluorescence}
  {Photoactivation} {Localization} {Microscopy}.
\newblock {\em Biophysical Journal}, 91(11):4258--4272, December 2006.

\bibitem{rust_sub-diffraction-limit_2006}
M.J. Rust, M. Bates, and X. Zhuang.
\newblock Sub-diffraction-limit imaging by stochastic optical reconstruction
  microscopy ({STORM}).
\newblock {\em Nature Methods}, 3(10):793--796, October 2006.

\bibitem{sage_quantitative_2015}
D. Sage, H. Kirshner, T. Pengo, N. Stuurman, J. Min, S.uliana
  Manley, and M. Unser.
\newblock Quantitative evaluation of software packages for single-molecule
  localization microscopy.
\newblock {\em Nature Methods}, 12(8):717--724, August 2015.

\bibitem{soubies_unified_2017}
E. Soubies, L. Blanc-Féraud, and G. Aubert.
\newblock A {Unified} {View} of {Exact} {Continuous} {Penalties} for l2-l0
  {Minimization}.
\newblock {\em SIAM Journal on Optimization}, 27(3), 2017.

\bibitem{soussen_bernoulli_2011}
C.~Soussen, J.~Idier, D.~Brie, and J.~Duan.
\newblock From {Bernoulli} - {Gaussian} {Deconvolution} to {Sparse}
  {Signal} {Restoration}.
\newblock {\em IEEE Transactions on Signal Processing}, 59(10):4572--4584,
  October 2011.

\bibitem{doi:10.1155/S168712000402009X}
S.~M. Stefanov.
\newblock Convex quadratic minimization subject to a linear constraint and box
  constraints.
\newblock {\em Applied Mathematics Research eXpress}, 2004(1):17--42, 2004.

\bibitem{yuan_sparsity_2016}
G. Yuan and B. Ghanem.
\newblock Sparsity {Constrained} {Minimization} via {Mathematical}
  {Programming} with {Equilibrium} {Constraints}.
\newblock {\em arXiv preprint arXiv:1608.04430} August 2016.


\end{thebibliography}
\end{document}